\documentstyle[12pt]{article}
\begin{document}
\baselineskip 24pt
\parskip 0.3cm
\parindent=0pt
\parskip 0.3cm
\parindent 0pt
\def\ref{\hangindent=1cm \hangafter=1 }

{\bf\centerline{Birth and early evolution of a planetary nebula}}
\vskip 1cm
 
{\centerline{M.\ Bobrowsky, K.C.\ Sahu, M.\ Parthasarathy, 
P.\ Garc\'{i}a-Lario}}
 
\vskip 8cm
{\centerline{To appear in {\it Nature}, 2 April 1998}}

\vskip 1cm
\vfill \eject

\vskip 0.3cm
{\bf\centerline{Birth and early evolution of a planetary nebula}}
\vskip 1cm

{\centerline{Matthew Bobrowsky$^1$, Kailash C. Sahu$^2$, M. Parthasarathy$^3$, Pedro Garc\'{i}a-Lario$^4$  }}

\vskip 1cm

{{$^1$Orbital Sciences Corporation,}
{7500 Greenway Center Drive, \#700, Greenbelt, MD 20770, USA}}

$^2$Space Telescope Science Institute, 3700 San Martin Drive, Baltimore, MD 21218, USA

$^3$Indian Institute of Astrophysics, Koramangala, Bangalore 560034

$^4$ISO Science Operations Centre, Villafranca del Castillo, Apartado de Correos 50727, 28080 Madrid, Spain

\vskip 1cm
\vfill \eject
{\bf The final expulsion of gas by a star as it forms a planetary 
nebula --- the ionized shell of gas often observed surrounding a 
young white dwarf ---
is one of the most poorly understood stages of stellar 
evolution$^{1,2}$.  Such nebulae form extremely rapidly (about 100 
years for the ionization) and so the formation process 
is inherently difficult to observe.  Particularly puzzling is 
how a spherical star can produce a highly
asymmetric nebula with collimated outflows.  Here we report
optical observations of the 
Stingray Nebula$^{3,4}$, which has become an ionized 
planetary nebula within the past few decades$^{5}$.  
We find that the collimated outflows are already evident, 
and we have identified the nebular structure that focuses the outflows.  
We have also found a companion star, reinforcing previous suspicions that 
binary companions play an important role in shaping planetary 
nebulae and changing the direction of successive outflows$^{7}$.}

Our narrow-band images (Fig.\ 1 and 2) show the compact nebula --- which 
we refer to as the `Stingray nebula' ---
that surrounds the central star of He3-1357 (also known as  
CPD-59$^\circ$6926, SAO 244567, and IRAS 17119-5926).  
In most of the images, the emission appears most strongly concentrated in 
an ellipse with its major axis subtending 1.$^{\prime\prime}$6 from 
northeast to southwest.  (This represents a correction from the 
orientation that we published previously$^{7}$.)  If the 
ellipse is actually a circular ring viewed obliquely, then our line of 
sight is inclined from the actual axis of symmetry by 56$^\circ$.  Above and 
below the ring are bubbles that are blown open at the poles.  In 
earlier images from the Wide Field and Planetary Camera 1 (WFPC1) --- 
obtained before the spherical aberration of the Hubble Space Telescope 
was corrected --- gas could just barely be seen escaping 
from these bubbles$^{7}$.  However, in the new WFPC2 images that we 
present here, the superior spatial resolution allows us to see that 
the escaping gas forms a complex system of collimated 
outflows.  (The nomenclature used for the various nebular features is 
shown in Fig. 3.)  In addition, the higher signal-to-noise ratio 
of the new images reveals 
that emission extends in all directions well outside of the major bubble 
and ring structures.  This is particularly evident in the low-ionization 
images.  In three of these images --- [OI] (shown in Fig.\ 2),
 [NII] and [SII]  (not shown, but similar to [OI]) ---  
the ring is much fainter and more mottled than in the other images.  
In most of the images (including the continuum image), 
the equatorial ring has dimensions 
1.$^{\prime\prime}$67 x 0.$^{\prime\prime}$92.  But in the 
[OI], [NII], and [SII] images, the minor axis is slightly longer, 
1.$^{\prime\prime}$0, presenting a visibly larger eccentricity of 
0.83 rather than 0.79 in the case of the other images.  In these 
same three images, the SE inner collimated outflow 
appears to be two separate, but closely situated, collimated 
outflows.  The overall shape of the nebula evokes the name 
`Stingray Nebula.'

The mass outflow that we observe is consistent with the blueshifted Si IV 
(1394-1403 \AA), C IV (1548-1551 \AA), and Al III (1855-1863 \AA) doublets 
that indicated the presence 
of a strong stellar wind$^{5}$. The low-ionization images 
show the strongest nebular emission coming from the collimated outflows  
close to the holes in the bubbles where gas is escaping.  
This is consistent with 
shock excitation caused by the wind encountering the shell and being 
focused by it.  Although converging flows have been 
observed previously$^{8}$, and hydrodynamic models have been used to simulate 
the formation of collimated outflows for almost two decades$^{9-14}$, 
this is the first observation that clearly shows the collimating 
{\it mechanism} of bipolar outflows.  Although theoretical work on 
bipolar outflows has often involved asymmetrical mass loss due to 
rapid stellar rotation or a toroidal magnetic field, or the presence
of disks (accretion disks or protostellar disks) to collimate the 
outflows, the latter process is not the mechanism observed in the 
Stingray Nebula.  The outflows in the Stingray are focused by the nebular 
bubbles which function like nozzles, with the gas leaving through the 
polar holes.  Bipolar structures have been 
seen both in proto-planetary nebulae (PPN) and in planetary nebulae (PN), 
but this is the only object that has been followed spectroscopically 
from its pre-ionized PPN stage$^{3,4}$ through ionization and 
recognition as a PN$^{5}$.  So it is especially 
interesting that the collimating mechanism is clearly visible in an 
object caught in this brief phase of its evolution.  
The H$\beta$ 
luminosity is similar to the intensity observed$^{7}$ 
in 1994 from which it was determined that the luminosity of the entire
nebular emission is $\sim$5000 times the solar luminosity$^{5}$ 
(assuming a distance of 5.6 kpc to the nebula$^{15,16}$). 

The dynamics in the Stingray Nebula are very complex, 
as evidenced by a number of additional features that are apparent in the 
HST images.  There are two axes of symmetry:  one defined by the 
bright inner ring, and one defined by the polar holes.  The new images 
presented here appear to show that the pairs of collimated, bipolar 
outflows outside 
of the polar holes have different orientations.  The position angles 
of stellar outflows are sometimes seen to vary monotonically with distance 
from the central star.  One example of this occurs in the young 
planetary nebula He3-1475 (refs. 8, 17, 18).  
This variation is often explained by monotonic precessional motion or a 
more chaotic ``wobbling,"$^{19,20}$ perhaps 
involving a companion object.  Depending on when the various 
outflows originate relative to the precessional cycle, the 
position angles of the collimated outflows can show some variation.  
In the Stingray Nebula, the position angles of successive 
outflow features appear to change with distance from the central star.
The inner collimated outflows, which  
have most recently emerged from the polar holes, are displaced in position 
angle from that of the polar holes by 3$^\circ$ in a counter-clockwise 
direction.  The position angles of the more extended (and therefore older) 
outer collimated outflows appear displaced by 
5$^\circ$ clockwise from the position angles of the inner collimated outflows.  
If the position of 
the polar holes controlled the direction of these outflows, then 
the precessional motion must involve the bubbles containing the 
polar holes.  

The WFPC1 images$^{7}$ did 
not have sufficient resolution to show a faint companion star embedded 
in the nebula.  But in the 
higher-resolution images presented here, a companion star is clearly 
visible.  The companion lies 0.$^{\prime\prime}$4 from the 
central star, at a position angle of 60$^\circ$, and, at 
magnitude V=17.0$\pm$0.2 as measured from the continuum filter centred 
at 6193 \AA, is 1.6 magnitudes fainter than the central star.  
(The central star's magnitude of V=15.4 is quite different from 
previously published magnitudes partly because the luminosity is 
dropping as seen from the comparison of ground-based observations$^{15,5}$ 
between 1985 and 1992, and as expected from the decreasing UV flux of 
the central star$^{5}$.  Perhaps more importantly, the older, ground-based 
observations could not resolve the nebula and therefore provided 
only a combined magnitude of the star plus the nebula.)  The 
magnitude determination at other wavelengths has higher
uncertainty  because of the strong nebular emission, but the
flux of the companion star is consistent with its being a late-type star
at the same distance as the nebula.  In considering whether 
this is truly a companion star, as opposed to a background star, we also made 
use of the IASG Galaxy Model$^{21,22}$ for the location of the Stingray Nebula 
(galactic longitude l=331$^\circ$, latitude b=$-12^\circ$).  At this location, 
the probability of finding this star so close (0.$^{\prime\prime}$4) 
to the central star is only $\sim$10$^{-4}$.  The probability is further 
reduced by a factor of $\sim$3
by the additional constraint on its distance as determined from its
color and luminosity.  Indeed, other than the central 
star, no stars this bright are visible anywhere else in the WFPC2 field 
of view.  These considerations leave little doubt that
the star is actually a binary companion. 

The detection of a binary companion is significant for several reasons.
First, the detection of a binary companion is consistent with
theoretical expectations.  For example, 
binary nuclei of planetary nebulae should
have a double peaked period distribution, and most 
should have a period between 30 and 300,000 years$^{6}$. (Below a 
critical orbital radius, the companion tends to spiral in 
during a ``common envelope" phase of mass loss. 
Above the critical orbital radius, mass-loss from the central star 
causes the orbital radius to increase as the gravitational potential 
decreases.)  The binary nucleus of
the Stingray Nebula, having a separation of about 2,200 AU, has a period of
about 100,000 years --- consistent with the theoretical expectations. 
Second, it has strong implications on the understanding of the early
evolution of bipolar structures in planetary nebulae.
There are essentially two ways to produce bipolar outflows:
one by a density contrast from the pole to the equator$^{23}$, and the
other by magnetic fields$^{24}$.  
In case of the close binaries, the development of a thick disk in the
common envelope phase causes a preferentially bipolar ejection of
material. In the case of a wide binary,
the companion helps by tidally spinning up the envelope$^{24}$.   
The spin up of the envelope by a companion also reduces the
minimum magnetic field required for magnetic effects to be
important. When the binary separation is large as seen here, 
the effect of the companion star would, at most, produce some 
small collimating effects$^{25}$ unless, as described above, its 
orbit has grown larger due to mass loss.  In the latter case, its effect 
on the shaping of the nebula may have been more profound.  
The detection of the binary companion in the case of
the Stingray nebula supports the proposed connection between binary 
central stars and asymmetrical planetary nebulae.
 
In addition to 
the overall axisymmetric structure that might be attributable to a 
companion star, there appears to be a distortion near the northern part 
of the equatorial ring.  (See ``spur" in Fig.\ 3.)  This is most easily visible 
in the H$\alpha$, H$\beta$, and [OIII] images, where gas from the ring 
appears to be drawn toward the companion star.  The image shows a number 
of similar wisps of gas throughout the nebula, so the presence of the spur 
near the companion star could be coincidental.  If, however, it is actually 
caused by the companion star, that would be notable because, although there has 
been a fair amount 
of theoretical work on how a companion star could distort a 
red-giant envelope, decreasing the pole-to-equator density ratio and 
thereby producing axial symmetry, the spur may be a case of 
{\it small-scale} distortions of a nebula caused by the binary companion.  

The electron temperature was derived using the [OIII] line intensity 
ratios (which are insensitive to the electron density).
The resulting temperature is of the order of $1.05 \times 10^{4}$ K, 
which is typical of planetary nebulae.  The electron density can be 
derived from the H$\alpha$ intensity$^{26,27}$ or the [SII] line 
ratio$^{5}$, which give a consistent value of $\sim$10$^{4}$ cm$^{-3}$.  
Given a distance of 5.6 kpc$^{15,16}$, the approximate radius 
of the nebula is 0.025 pc.  A density of 10$^{4}$~cm$^{-3}$ would then imply 
a nebular mass of 0.015 times the solar mass (0.015 M$_\odot$). There is 
no strong abundance variation 
in the nebula as seen from the ratio of the line intensities.

A spectrum taken with the International Ultraviolet Explorer (IUE) in 1988 
showed$^{5}$ a strong P-Cygni profile 
(an emission feature with a blue-shifted absorption component) 
suggesting the presence of a fast wind with terminal velocity 
of --3,060 km s$^{-1}$. The C IV intensity has since decreased monotonically
and had fallen below the detection limit by 1994.$^{28,29}$ Our 
spectrum, taken in 1997,
does not show any C IV emission, which confirms that the fast wind has stopped.
(The outer shell is the result of the past, slow, red-giant wind.)   
Because the appearance of multiple 
collimated outflows implies that the stellar wind is episodic, it is 
possible that the stellar wind will appear some time again. However,  
the fact that the UV Flux has also decreased by 
a factor of 3 within the past eight years$^{28}$, combined with the C IV line 
decreasing to an undetectable point now, 
suggests that the central star has, for now, become devoid of nearby 
circumstellar material.  At the same time, the
ionization state of the nebula has increased considerably, which
suggests that the temperature of the central star is higher. 

The simultaneous drop in the UV flux and the increase in temperature 
at this stage imply a drop in the total luminosity of the central star.
This places the central star of the Stingray nebula at
the ``knee" of the Hertzsprung--Russell diagram$^{30}$, just before the star
evolves towards lower temperatures and lower luminosities.  However, 
these observations reveal some inadequacy in the current theoretical 
understanding of how these stars evolve.  For the central star to 
evolve as rapidly as observed here, 
a stellar mass of 0.8M$_\odot$ or more is necessary.  However, 
the luminosity of the central star indicates a core mass of 0.6M$_\odot$ 
or less, for which the evolution is expected to be much slower than 
observed$^{31}$.  
The mass loss in the post-AGB (asymptotic giant branch) may be the 
cause of the unexpectedly 
rapid evolution and may also cause the differences between observations 
and theory. 
The observations clearly show that the central star has become a 
planetary nebula only within the past few decades$^{5}$, and is 
rapidly evolving to become a DA white dwarf$^{~29}$.  Although much 
work on protoplanetary nebulae and other bipolar objects has been 
done in an attempt to understand the origin of the axial symmetry 
and collimated outflows, the Stingray Nebula shows how far the nebular 
structure can develop by the time the nebula becomes ionized.  We 
believe that no other planetary nebula in this phase of its evolution 
has been previously identified. 

\vskip 1cm
{\bf References:}
\vskip 0.1cm

\ref 1. Kwok, S. Transition from Red Giants to Planetary Nebulae, in 
{\it Late Stages of Stellar Evolution}, eds S. Kwok \& 
S.R. Pottasch., 321--335 (Kluwer Academic Press, Dordrecht, 1987).\par 

\ref 2. Maddox, J. Making sense of dwarf-star evolution. Nature, {\bf 376}, 
15 (1995).

\ref 3. Henize, K.G. Observations of Southern Planetary Nebulae.  
Astrophys.\ J.\ Suppl.\ Ser., {\bf 14}, 125--153 (1967).\par 

\ref 4. Henize, K.G. Observations of Southern Emission-Line Stars. 
Astrophys.\ J.\ Suppl.\ Ser., {\bf 30}, 491--550 (1976).\par 

\ref 5. Parthasarathy, M. et al. SAO 244567: a post-AGB star which has 
turned into a planetary nebula within the last 40 years.  Astron.\ 
Astrophys., {\bf 267}, L19--L22 (1993).\par 

\ref 6. Youngelson, L.R., Tutukov, A.V., Livio, M.,
The formation of binary and single nuclei of planetary nebulae,
Astrophys.\ J., {\bf 418}, 794-803 (1993).

\ref 7.  Bobrowsky, M. Narrowband HST Imagery of the Young Planetary 
Nebula Henize 1357.  Astrophys.\ J., {\bf 426}, L47--L50 (1994).

\ref 8.  Borkowski, K.J. et al. Collimation of Astrophysical Jets:  
The Proto-Planetary Nebula He3-1475.  Astrophys.\ J., {\bf 482}, 
L97--L100 (1997).

\ref 9.  Cant\'{o}, J. A Stellar Wind Model for Herbig-Haro Objects.  
Astron.\ Astrophys., {\bf 86}, 327--338 (1980).

\ref 10. Morris, M. Bipolar Symmetry in the Mass Outflows of Stars in 
Transition; in {\it From Miras to Planetary Nebulae}, eds 
Mennessier, M.O. \& Omont, A., 520--535 (Editions Fronti\`{e}res, 
Gif-sur-Yvette, 1990).

\ref 11. Soker, N. \& Livio, M. Disks and Jets in Planetary Nebulae. 
Astrophys.\ J., {\bf 421}, 219--224 (1994).

\ref 12. Mellema, G.,  Astrophys.\ \& Space Sci., {\bf 245}, 239--253 (1996).

\ref 13. Frank, A. et al. A Mechanism for the Production of Jets and 
Ansae in Planetary Nebulae.  Astrophys.\ J. Lett., {\bf 471}, L53--L56 (1996).

\ref 14.  Frank, A. \& Mellema, G. Hydrodynamic Collimation of YSO Jets. in 
{\it Herbig-Haro Flows and the Birth of Low Mass Stars}; eds Reipurth, B. 
\& Bertout, C., 291--302 (Kluwer Academic Press, Dordrecht, 1997).

\ref 15. Kozok, J.R. Photometric observations of emission B-stars 
in the southern Milky Way.  Astron.\ Astrophys.\ Suppl.\ Ser., 
{\bf 61}, 387--405 (1985).

\ref 16. Kozok, J.R. Distances, reddening, and distribution of emission 
B-stars in the galactic centre region $|l| \leq 45^\circ$.  Astron. 
Astrophys.\ Suppl.\ Ser., {\bf 62}, 7--16 (1985).

\ref 17. Bobrowsky, M. et al. He3-1475 and its Jets.  Astrophys.\ J., 
{\bf 446}, L89--L92 (1995).

\ref 18. Riera, A. et al. IRAS 17423--1755 [He3-1475]: a massive post-AGB 
star evolving into the planetary nebula stage?  Astron.\ Astrophys., 
{\bf 302}, 137--153 (1995).

\ref 19. Livio, M. \& Pringle, J.E. The Formation of Point-Symmetric 
Nebulae.  Astrophys.\ J., {\bf 465}, L55--L56 (1996). 

\ref 20. Livio, M. \& Pringle, J.E. Wobbling Accretion Disks, Jets, and 
Point-Symmetric Nebulae.  Astrophys.\ J., {\bf 486}, 835--839 (1997).

\ref 21.  Ratnatunga, K.U., Bahcall, J.N., \& Casertano, S.,
 Kinematic Modeling of the Galaxy I.~The Yale Bright Star Catalog,
 Astrophys.\ J., {\bf 339}, 106--125 (1989).

\ref 22.  Casertano, S., Ratnatunga, K.U. \& Bahcall, J.N.,
 Kinematic Modeling of the Galaxy.  II.~Two Samples of High Proper Motion Stars,
 Astrophys.\ J., {\bf 357}, 435--452 (1990).

\ref 23. Icke, V., Balick, B. \& Frank, A., The hydrodynamics of 
aspherical planetary nebulae II. Numerical modelling of the early 
evolution, Astron.\ Astrophys., {\bf 253}, 224-243 (1992).

\ref 24. Livio, M., HST Observations of Nebular Morphologies and their 
Implications, Space Sci. Rev., in press.

\ref 25. Kolesnik, I.G., \& Pilyugin, L.S., The influence of
a binary nucleus on the structure of planetary nebulae, Sov. Astron.,
30, 169-173 (1986).

\ref 26. Sahu, K.C. \& Desai, J.N. Kinematic structure of NGC 3132: the 
planetary nebula with a binary nucleus.  Astron.\ Astrophys.,  
{\bf 161}, 357--362 (1986).

\ref 27. Parthasarathy, M. et al. Fading and variations in the spectrum 
of the central star of the very young planetary nebula SAO 244567 
(Hen 1357).  Astron.\ Astrophys., {\bf 300}, L25--L28 (1995).

\ref 28. Pottasch, S.R., {\it Planetary Nebulae}, 51--70 (D. Reidel Publ.
Co., 1983). 

\ref 29.  Feibelman, W.A. Recent Changes in the IUE Spectrum of the Very 
Young Planetary Nebula He3-1357 (=SAO 244567).  Astrophys.\ J., {\bf 443}, 
245--248 (1995).

\ref 30.  Bl\"ocker, T. \& Sch\"onberner, D. Stellar evolution of low and 
intermediate-mass stars. III. An application of evolutionary post-AGB 
models: the variable central star FG Sagittae. Astron.\ Astrophys., 
{\bf 324}, 991--997 (1997).

\ref 31.  Bl\"ocker, T. \& Sch\"onberner, D. On the fading of massive 
AGB remnants. Astron.\ Astrophys., {\bf 240}, L11--L14 (1990).

\vskip 1cm
Acknowledgements: We thank T. Heckman for suggestions, M. Livio for 
discussions, and A. Frank for comments.  This work is based on observations 
with the NASA/ESA 
Hubble Space Telescope, obtained at the Space Telescope Science Institute, 
which is operated by the Association of Universities for Research in 
Astronomy, Inc., under NASA contract.

Correspondence and requests for materials should be addressed to M.B. (e-mail:
mattb@cta.com).
\vfill \eject
 
{\bf Figure Captions:}

{\bf Fig.\ 1.}  A Hubble Space Telescope composite image of the 
Stingray Nebula reveals complicated streams of outflowing gas, and 
a companion star visible within the nebula.  Red, green, and blue 
indicate light emitted at wavelengths of 6583 \AA, 5007 \AA, and 
4861 \AA, from nitrogen, oxygen, and hydrogen, respectively.

{\bf Fig.\ 2.}  Hubble Space Telescope narrow-band images of He3-1357, 
the Stingray nebula.  The images were obtained in four 
emission lines (H$\alpha$, He I 5876 \AA, [O I] 6300 \AA, and 
[O III] 5007 \AA), a continuum region of the spectrum centred at 6193 \AA, 
and the ratio of [OIII]5007\AA\ to H$\beta$.  Observations were made 
in March 1996 with the 
WFPC2 using the narrow-band filters 
F487N, F502N, F588N, F630N, and F656N.  The continuum 
exposure was made using filter FQCH4N15 which has a passband of
$\sim$44\AA\ centred at 6193\AA.  Although 
not shown in this figure, images were also acquired with filters 
F658N ([NII] 6583 \AA) and F673N ([SII] 6731 \AA) which show the 
nebula with an appearance almost identical to the [O I] image shown here.  
The routine scientific data processing included bias, preflash, dark, 
and flat-field corrections.  After removing bad pixels resulting 
from cosmic rays, both images from the same filter were aligned 
and averaged.  The ratio of the [OIII] 5007\AA\ to H$\beta$ 
emission (bottom right panel) varies across the nebula.  Relative 
to H$\beta$, the [OIII] emission is lowest ([OIII]/H$\beta \sim 2$) 
near the central star, and is also low ([OIII]/H$\beta \sim 4$) 
in the denser regions of the bubbles where the gas has been most compressed.  
The low [OIII]/H$\beta$ ratio in these areas is presumably due to 
collisional de-excitation occurring where the density is higher.  
The [OIII]/H$\beta$ ratio is highest (up to 20) in regions of 
intermediate density, mostly towards the outer parts of the nebula. 

{\bf Fig.\ 3.} An [O III]5007\AA\ image with various morphological features 
indicated.  See text for details.

\end{document}